\documentclass[a4paper,referee]{aa}
\usepackage{subfigure}
\usepackage{natbib}
\usepackage{verbatim}
\usepackage{graphicx}
\usepackage{amssymb}  
\usepackage{amsmath}
\newcommand{\new}{}

\begin{document}

\title{Guided flows in coronal magnetic flux tubes}
 
\author{A. Petralia\inst{1} \and F. Reale\inst{2,1} \and P.Testa\inst{3}}

\institute{INAF-Osservatorio Astronomico di Palermo, Piazza del Parlamento 1, 90134 Palermo, Italy
\and Dipartimento di Fisica \& Chimica, Universit\`a di Palermo, Piazza del Parlamento 1, 90134 Palermo, Italy
\and Smithsonian Astrophysical Observatory, 60 Garden Street, MS 58, Cambridge, MA 02138, USA}

\date{Rec;Acc}

\abstract
{There is evidence for coronal plasma flows to break down into fragments and to be laminar.}{We investigate this effect by modeling flows confined along magnetic channels.} 
{We consider a full MHD model of a solar atmosphere box with a dipole magnetic field. We compare the propagation of a cylindrical flow perfectly aligned to the field to that of another one with a slight misalignment. We assume a flow speed of 200 km/s, and an ambient magnetic field of 30 G.}
{We find that while the aligned flow maintains its cylindrical symmetry while it travels along the magnetic tube, the misaligned one is rapidly squashed on one side, becoming laminar and eventually fragmented because of the interaction and backreaction of the magnetic field. This model could explain an observation of erupted fragments that fall back as thin and elongated strands and end up onto the solar surface in a hedge-like configuration, made by the {\it Atmospheric Imaging Assembly} on board the {\it Solar Dynamics Observatory}.}
{The initial alignment of plasma flow plays an important role in determining the possible laminar structure and fragmentation of flows while they travel along magnetic channels.}

\keywords{magnetohydrodynamics (MHD) - Sun activity - Sun:corona}

\maketitle

\section{Introduction}

Plasma flows are ubiquitous in the solar corona. They can be generated by many different processes characterized by the interaction between the strong and complex magnetic field of the corona and its plasma.  
The interaction can be relatively weak and it can lead to confined flows, e.g., spicules \citep{Beckers68,Depontetal07}, siphon flows \citep{Ruedetal92,Breketal97,Maretal92,Winetal01,Winetal02,Teretal04}, coronal rain \citep{Parker1953,Field1965,Antetal2012} or it can be strong leading to violent ejection of plasma outside of the Sun as in Coronal Mass Ejections \citep[CMEs,][]{Chen2011,Webb2012}.

In an impressive eruption event that occurred on 7 June 2011, the ejected material fell back along parabolic trajectories and impacted the solar surface \citep{Carletal2014,vanDrieetal2014,Innetal2012}, producing brightenings \citep{Reaetal2013}. In this case, the dynamics is strongly influenced by the intensity and the complexity of the magnetic field close to the impact region. In particular, impacts where the magnetic field is weak brighten because of the shock heating of their outer shells \citep{Reaetal2013,Reaetal2014}, other plasma blobs are channelled by a strong magnetic field and the shocks brighten the whole channel ahead \citep{Petetal2016}. \cite{Petetal2017} show that the misalignment between the velocity of the blobs and the magnetic field lines determines a significant disruption of the blobs.

The misalignment between plasma flows and the ambient magnetic field can therefore be important. Magnetohydrodynamic modeling of the solar chromosphere up to the corona has shown that chromospheric flows can be not well aligned with the magnetic field \citep{Maretal2016}. As we will show, there is evidence for this misalignment also from the observations.

In this work, we address specifically the possible role of the initial field alignment or misalignment of a persistent outflow ejected in a magnetized atmosphere. 
Our approach is to model flows which are pushed upward from the chromosphere along closed magnetic flux tubes in the corona by means of 3D-MHD simulations able to capture the relevant physics behind the process and to take into account other important effects, such as the natural expansion of the magnetic channel cross section with the height and its effect on the dynamics, or possible transversal motions of the flow \citep[eg.][]{Petetal2016}. We first shows an observed sample case of laminar fragmented flow in Section~\ref{sec:obs}. The model is described in Section~\ref{sec:model}. The simulations and the results are presented in Section~\ref{sec:results}, and we discuss them in Section~\ref{sec:disc}.

\section{An observed sample case}
\label{sec:obs}

  	\begin{figure*}[!th]
	\centering
	\includegraphics[width=12cm]{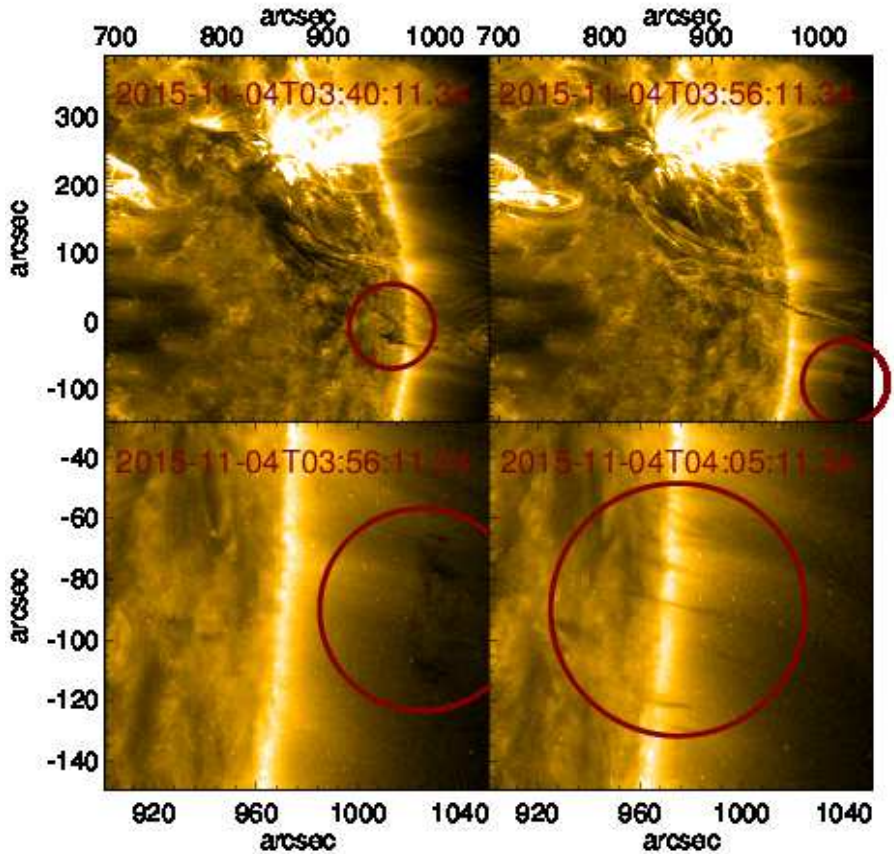}
	\caption{Evolution of falling material observed in the 171 \AA band of SDO/AIA on 4 November 2015 at the labelled times (see the associated movie).}
	\label{fig:obs}
	\end{figure*}

After a solar eruption on 4 November 2015, SDO/AIA observed dense fragments to fall back onto the solar surface. During the fall they elongated, spread and then impacted the solar surface in a row with an hedge-like configuration, as shown in Figure~\ref{fig:obs} and in the associated movie. This evolution might appear not obvious but our present model may explain it. 
The fragments are observed in absorption and this allows us to estimate their density and temperature according to the method in \citep{LaneRea2013}. We obtain a density of  $\sim 1.5 \times 10^{10}$~cm$^{-3}$ and a temperature of $\sim 4 \times 10^4$~K. We also estimate the impact velocity perpendicular to the solar surface by assuming a parabolic motion. We measure the vertical distance ($H$) covered from the apex of the trajectory to the impact region and we obtained for the final velocity $v=\sqrt{2g_{\sun}H} \sim200$~km/s, where $g_{\sun}$ is the solar gravity. 
The observed evolution and destiny of the erupted fragments suggest this as an example of an outflow that is initially not perfectly aligned to the magnetic field coming out of the solar surface, as illustrated in the following Sections.

\section{The Model}
\label{sec:model}

We use the model presented in \cite{Petetal2016} and the numerical code PLUTO \citep{MignZan12} to describe the dynamics of a persistent flow injected upwards and confined inside a closed coronal magnetic flux tube. This recalls closely the dynamics of siphon flows. We consider two slightly different flow directions, one perfectly aligned to the field and the other slightly inclined with respect to the field lines. In the following, we will call the former `Aligned Flow' and the latter `Misaligned Flow'.

The ambient atmosphere is made by a stratified million-degree corona linked to a much denser chromosphere through a steep transition region. The corona is a hydrostatic atmosphere \citep{RTV1978,Seretal1981} that extends upwards for $10^{10}$cm. The chromosphere below is hydrostatic and isothermal at $10^4$K with a density at the base of $\sim 10^{16}$cm$^{-3}$. The atmosphere is made plane-parallel along the vertical direction ($Z$).

\begin{figure*}[!ht]
	\centering
	\includegraphics[scale=0.7]{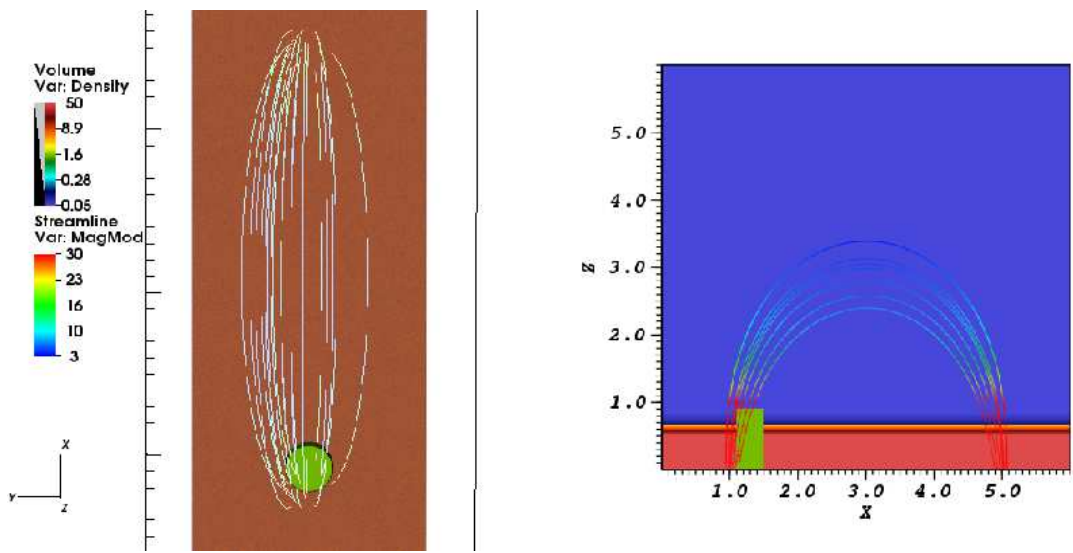}
	\caption{Initial model conditions for the misaligned flow. Rendering of the density (units of $2\times10^{10}$cm$^{-3}$, logarithmic scale) viewed from above, i.e., the line of sight is aligned with the Z axis (left panel), and from the side, i.e. on an $XZ$ plane (right panel) \new{, in the computational domain (axis in units of $10^9$~cm)}. A bundle of magnetic field lines \new{(Gauss)} is also shown.}
	\label{fig:inicond_flow_rend}
	\end{figure*}

The closed magnetic loop has been obtained by setting a dipole magnetic field centred on and parallel to the solar surface ($XY$). The field intensity is $\sim30$G at the top of the chromosphere and rapidly decreasing with the height (Fig.~\ref{fig:inicond_flow_rend}). The field confines the flow but can be perturbed by it; indeed the ratio between the ram pressure carried by the flow and the magnetic pressure is $\frac{\rho v^2}{B^2/8\pi} \sim 0.7$ at the top of the transition region. This ratio may be expected in an active region loop.

We consider the same ambient atmosphere as the `Dense Model' described in \cite{Petetal2016}, that -- combined to the field intensity -- gives an Alfv\'en speed of $\sim 2000$ km/s at the top of the chromosphere close to the footpoint of the loop.

\begin{figure*}[!ht]
	\centering
	\includegraphics[scale=0.5]{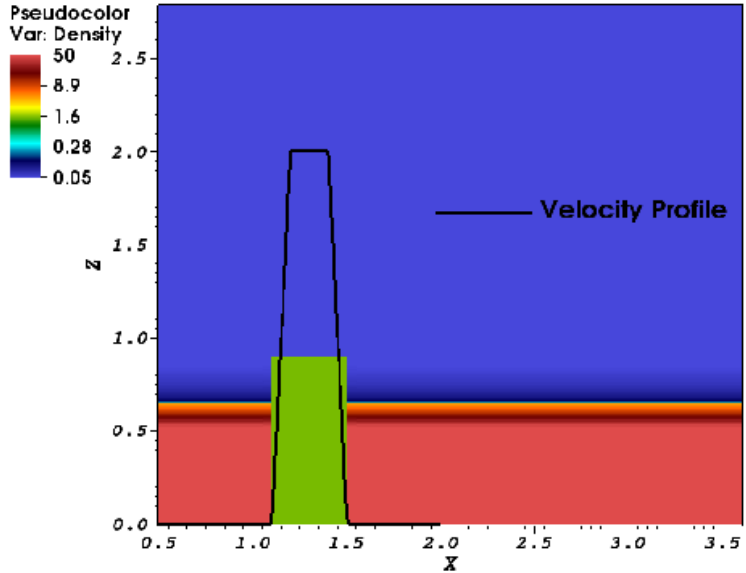}
	\caption{Same as Fig.~\ref{fig:inicond_flow_rend} (right) but including the initial spatial profile of the  velocity \new{across the flow (black line) in units of $100$~km/s.} }
	\label{fig:2upflow}
	\end{figure*}

\new{Figures~\ref{fig:inicond_flow_rend} and \ref{fig:2upflow} show our initial conditions.} The flow is injected upwards from a localized area of the chromosphere \new{and is continuously fed from the lower domain boundary, i.e., the injection is set as a boundary condition inside the flow area at the base of the chromosphere. The flow area is a circle with a radius $R_F = 2\times10^8$cm. The flow is present since the beginning inside the domain above the lower boundary, as shown in Fig.~\ref{fig:inicond_flow_rend}, with the shape of a cylinder with a vertical length of $L_F = 10^9$cm. Its initial density and temperature are uniformly $3\times10^{10}$~cm$^{-3}$ and $4\times10^4$~K, respectively.} Since the thickness of the chromosphere is $\approx 7 \times 10^8$~cm, the tip of the jet protrudes in the corona already since the beginning.

The misaligned flow is injected at a distance $X=1.3\times10^9$~cm from the left boundary where the magnetic field lines rapidly curve and leads to a misalignment of the flow. The flow propagates in the quasi semi-circular flux tube that has a distance between the footpoints $\sim 4\times10^9$~cm and a height of $2.5\times10^9$~cm above the chromosphere.

\new{As shown in Fig.\ref{fig:2upflow},} the velocity of the flow is uniform along the $Z$ direction $v_Z=200$~km/s inside an inner circular area with radius $R=10^8$~cm, and then linearly decreasing to zero in an outer shell to the cylinder boundary $R=2\times10^8$~cm. This initial velocity is such that the flow has enough kinetic energy to reach and surpass the apex of the closed flux tube. 

The aligned flow is injected instead at a smaller distance $X=0.9\times10^9$~cm from the left boundary, where the magnetic field lines are nearly open and thus vertical because closer to the magnetic pole and aligned with the vertical direction of the flow. This flow will travel to a much larger height along the flux tube (Fig.~\ref{fig:2upflow}). 

The computational box is three-dimensional and Cartesian ($X$, $Y$, $Z$) and extends over $6\times10^9$~cm in the $X$ direction, $1\times10^9$~cm in the $Y$ direction and $6\times10^9$~cm in the $Z$ direction, that is perpendicular to the solar surface. The mesh of the 3D domain is adaptively refined to strong gradients of density. The roughest level has $120\times20\times120$ number of cells and it is refined up to 3 levels. Each level refines locally the mesh by a factor of two, giving a cell size of $\sim60$~km at the highest level of refinement. The geometry of the system is symmetric with respect to the $Y=0$ plane, allowing us to simulate only half of the domain.

\new{Therefore, at the $Y=0$ plane boundary we impose reflecting conditions. 
At all the other boundaries the magnetic field is fixed. For the other physical variables, we impose outflow conditions at all boundaries except for the plane at the base of the chromosphere, i.e. at $Z=0$. There, we have fixed values and zero velocity outside the flow area, and inflow condition inside the flow area, i.e. a fixed velocity with the same radial dependence as in the inner domain. Thus, the flow velocity at the boundary is also constant in time}.
 
\section{Results}
\label{sec:results}
	
\subsection{The aligned flow}

\begin{figure*}[!th]
	\centering
	\resizebox{\textwidth}{!}
	{
	\includegraphics[width=0.39\columnwidth ,angle=-90]{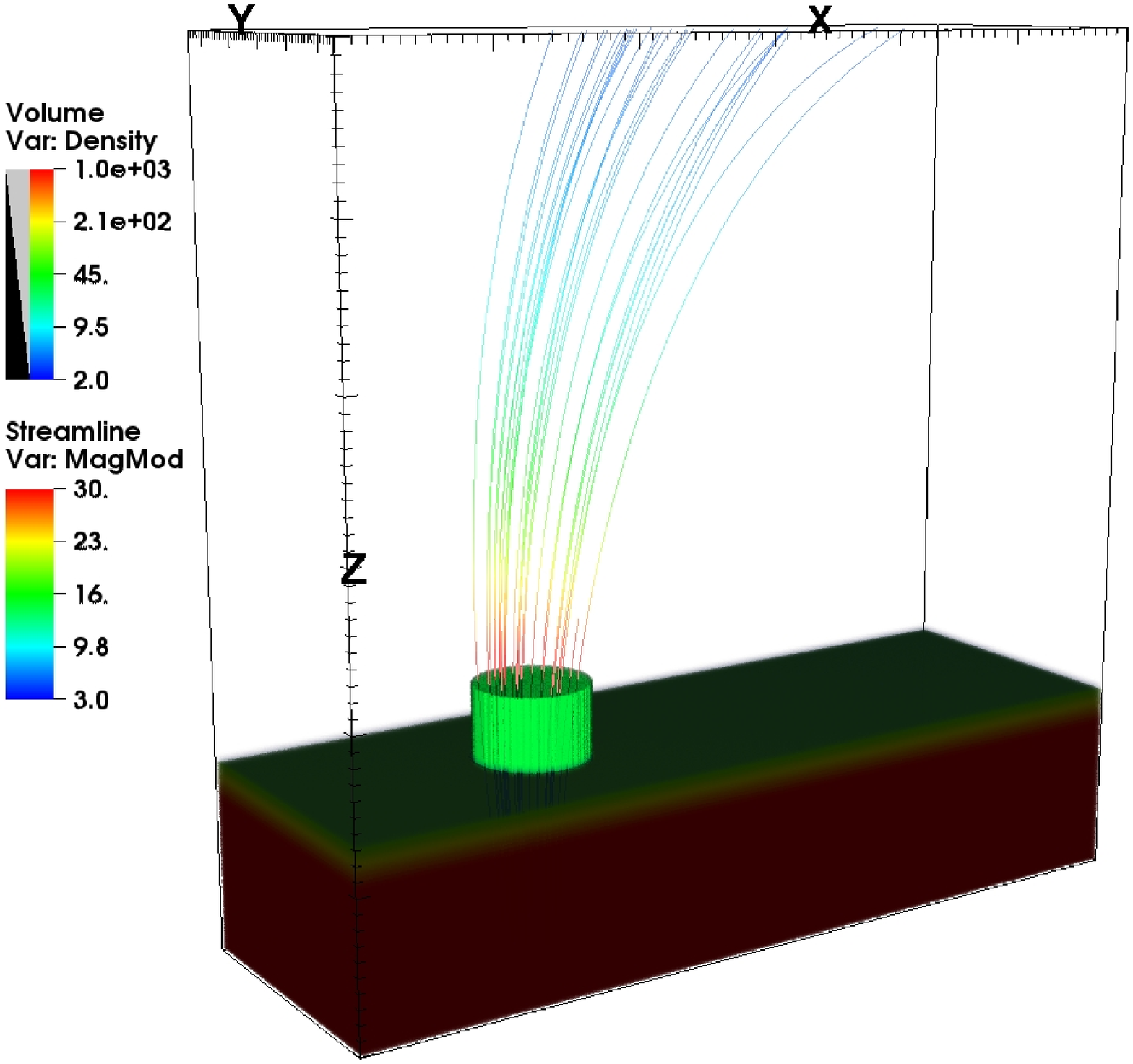}  
	\includegraphics[width=0.39\columnwidth ,angle=-90]{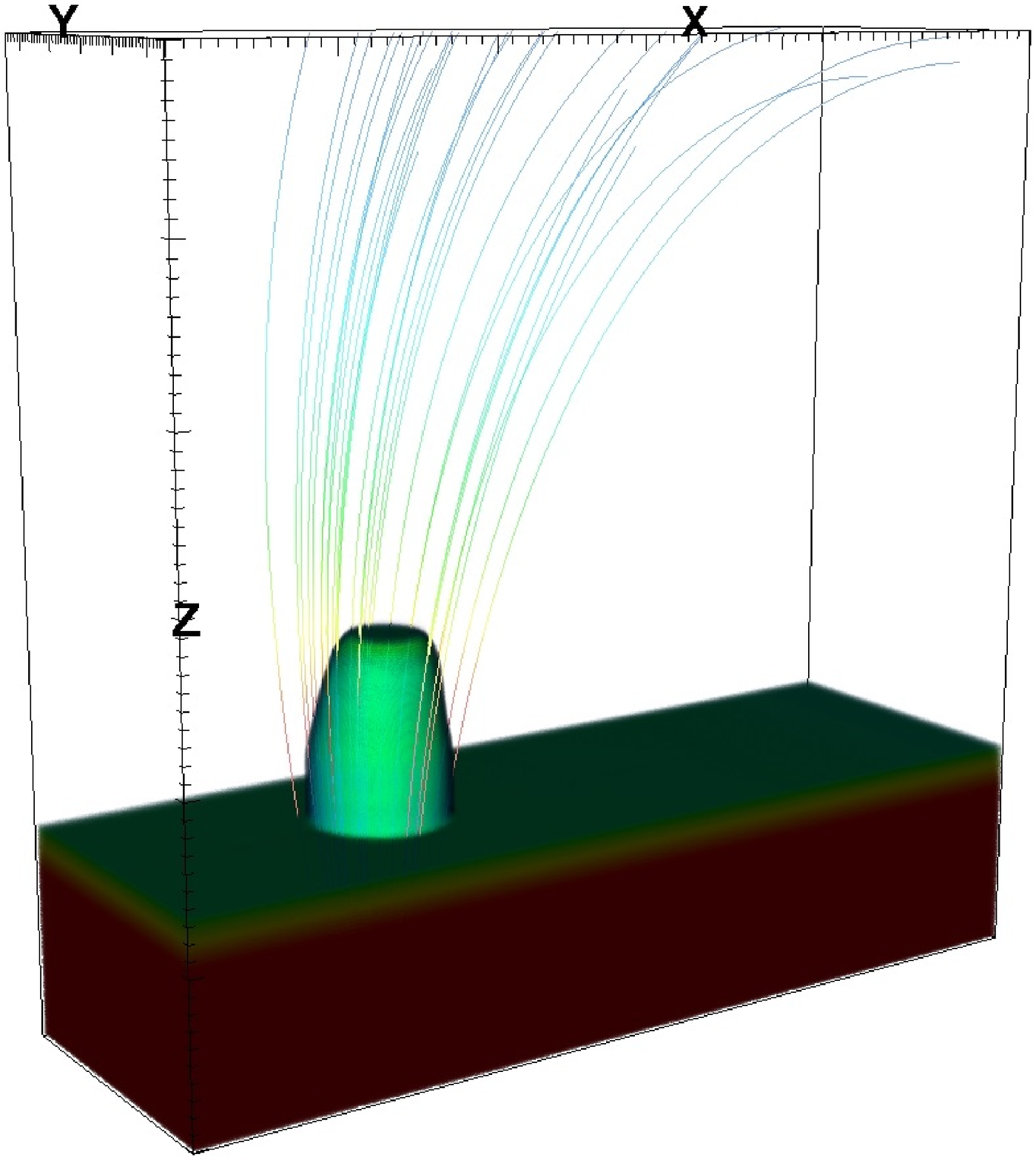}
	}
	\resizebox{\textwidth}{!}
	{
	\includegraphics[width=0.39\columnwidth ,angle=-90]{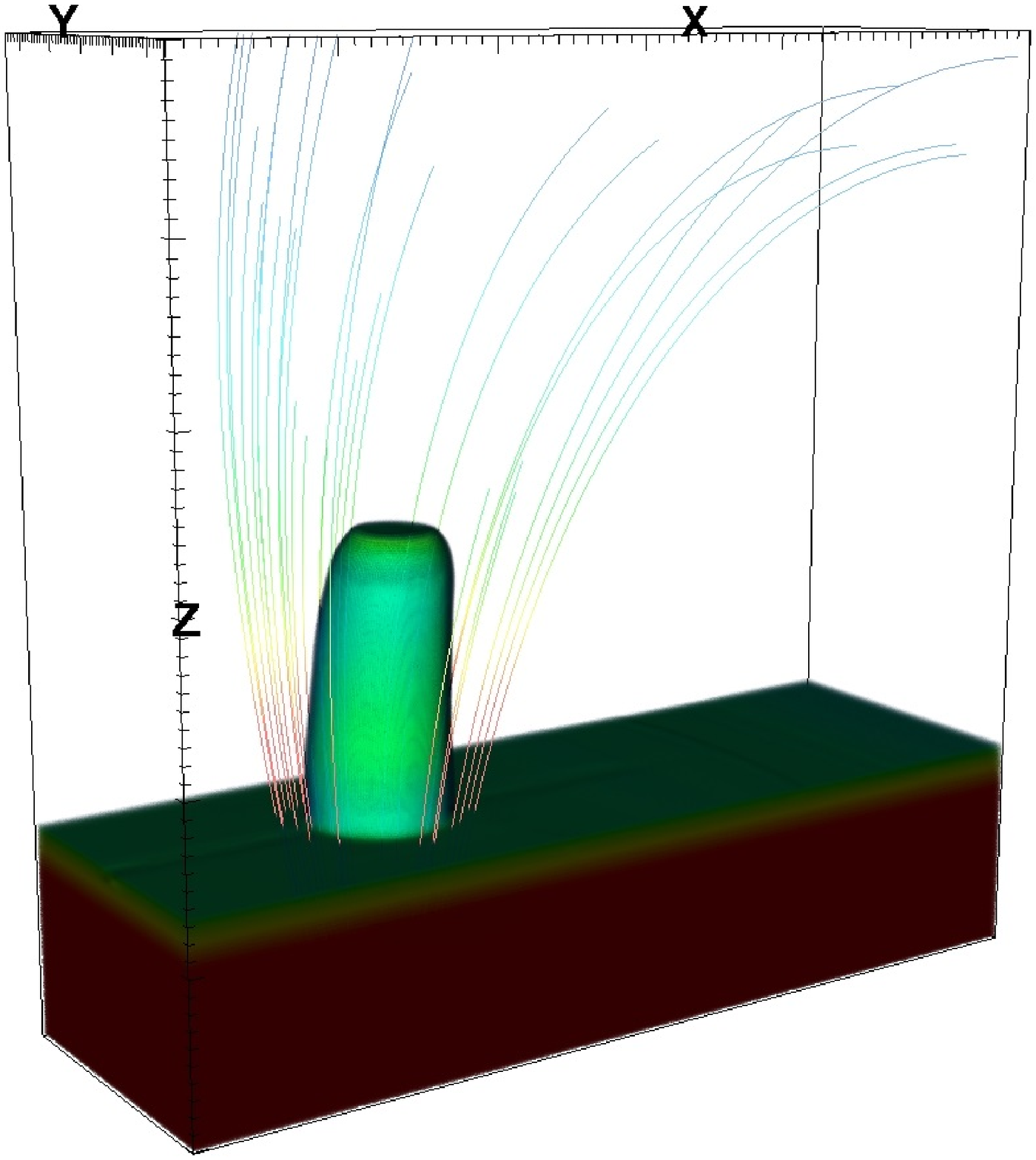}
	\includegraphics[width=0.39\columnwidth ,angle=-90]{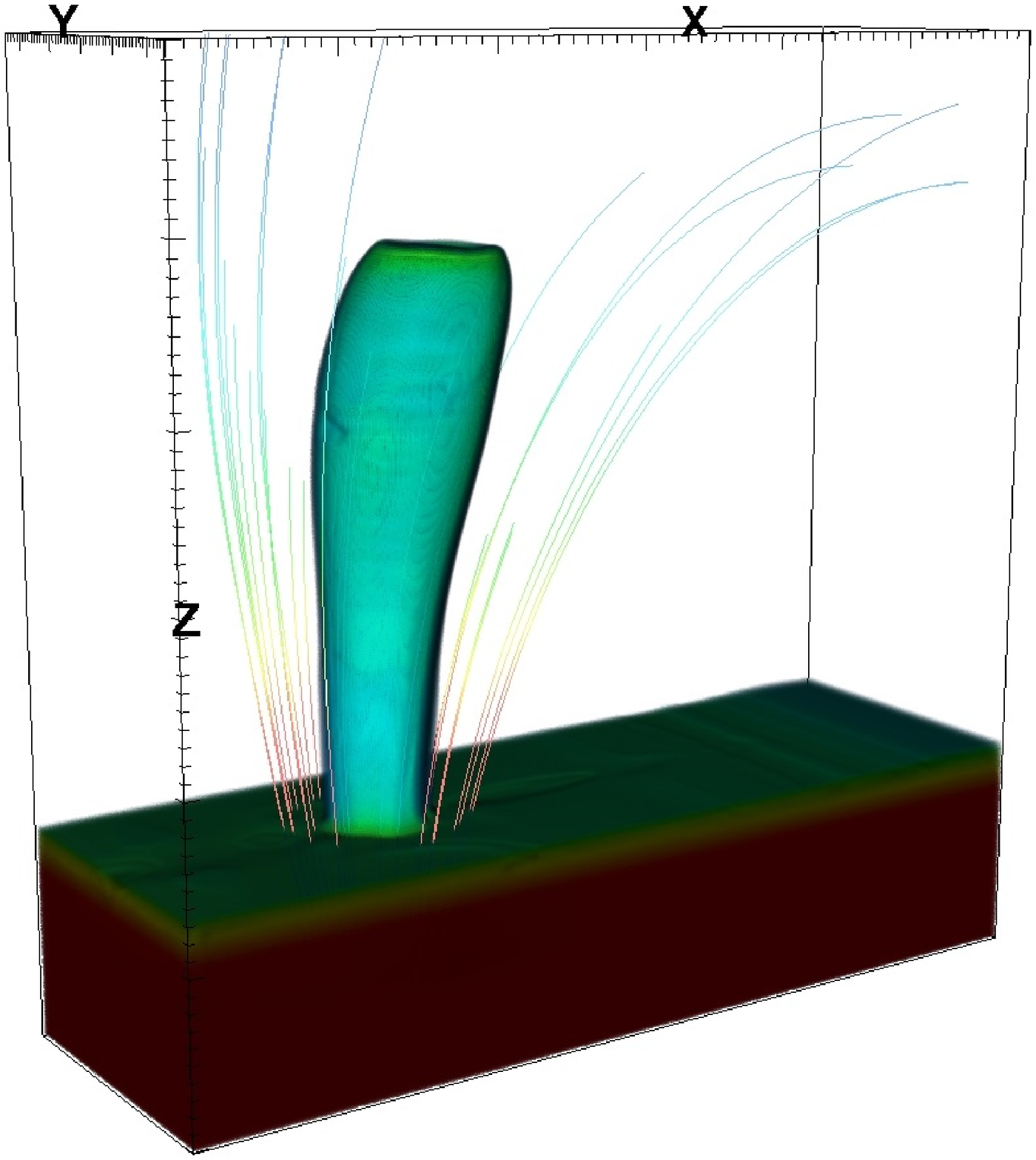}  
	}
	\caption{Simulation of the aligned flow: Rendering of the density ($10^{9}$cm$^{-3}$, logarithmic scale) at times t=0, 20, 40, 100s in which a bundle of magnetic field lines are shown \new{(see the associated movie)}.}
	\label{fig:evol3D_alig}
	\end{figure*}

Figure~\ref{fig:evol3D_alig} shows four snapshots of the evolution of the aligned flow. The flow propagates in a flux tube that is almost vertical at the injection site; therefore, it reaches a maximum height and eventually falls back onto the chromosphere. 

\begin{figure*}[!th]
	\centering
	\includegraphics[width=14cm]{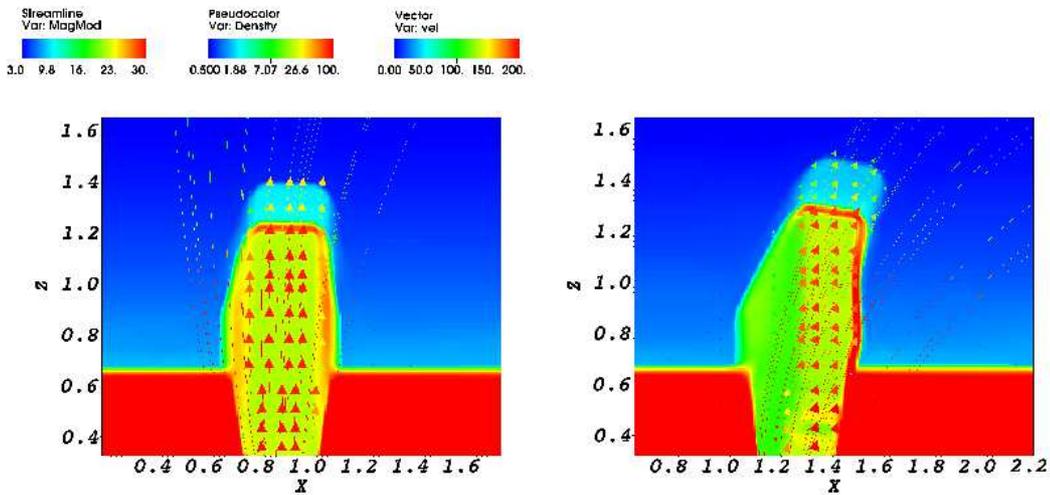}
	\caption{Density maps (units of $10^{9}$cm$^{-3}$, logarithmic scale) on an $XZ$ cross-section across the center of the domain  at time t=20~s, for the aligned (left) and for the misaligned flow (right). A bundle of magnetic field lines \new{(Gauss)} and the velocity field \new{(km/s)} are also shown.}
	\label{fig:comp_vert}
	\end{figure*}

The jet is initially uniform along the vertical direction, while the surrounding atmosphere is not. So the flow is not in pressure equilibrium with the ambient atmosphere:  
in particular, below the transition region (not shown \new{in Fig.~\ref{fig:evol3D_alig}, see Fig.\ref{fig:comp_vert}}), at the lower boundary, the ambient chromosphere is very dense, and the thermal pressure ($p_{chrom}>10^{4}$ dyn cm$^{-2}$) is much larger than inside the flow ($p_{flow}\sim 10^{-1}$ dyn cm$^{-2}$, $p_{mag}\sim 4\times10^{2}$ dyn cm$^{-2}$). Therefore, at its origin, both the flow and the magnetic field inside it are rapidly squeezed by the outside pressure and the flow is \new{de facto quenched} after $\sim40$~s of evolution. \new{This might not be entirely realistic, although we may expect that such rapid flows do not have enough time to reach pressure equilibrium while crossing the chromosphere. However, the initial supply is large enough to sustain the flow along the magnetic channel. This applies also to the case of misaligned flow described in Section~\ref{subsec:shockdyn}. }

Since the initial speed of the flow ($v_{flow}=200$km/s) exceeds the (isothermal) sound speed ($c_s= \sqrt{\gamma\frac{2K_bT}{\mu m_p}} \sim130$~km/s, with $T\sim8\times10^5$~K),  \new{a shock front is generated. This is a slow-mode shock and moves ahead of the flow (see the pale blue cocoon in Fig.~\ref{fig:comp_vert}) along the magnetic field lines. Its initial compression ratio of the density is $\rho_{post-shock}/\rho_{pre-shock} \approx 2.8$, and the temperature grows by $\sim 15\%$ in the post-shock. The overall propagation is similar to \cite{Petetal2016}. }

\begin{figure*}[!th]
	\centering
	\includegraphics[width=7cm, angle=-90]{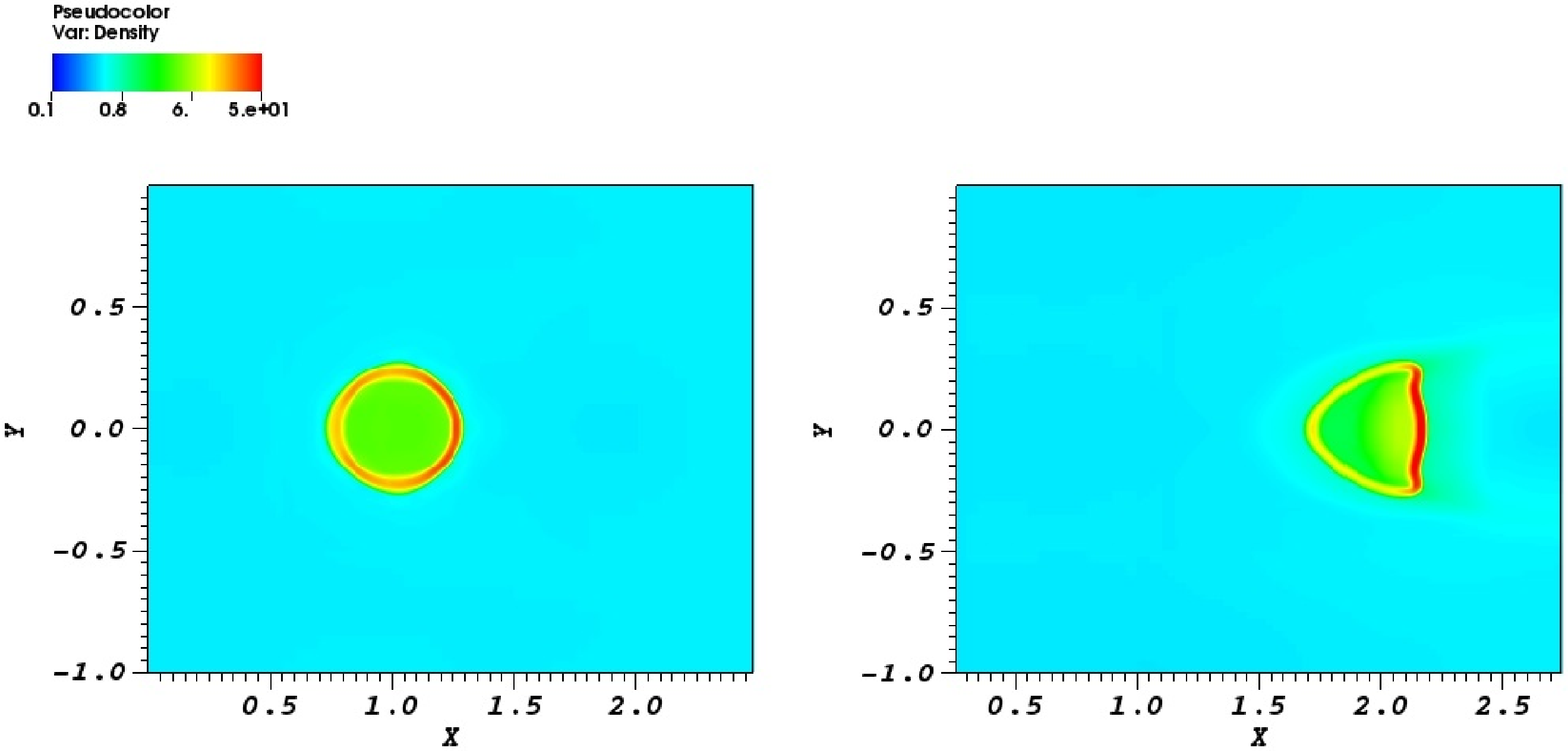}
	\caption{Density maps (units of $10^{9}$cm$^{-3}$, logarithmic scale) on a $XY$ cross-section at $Z=2\times10^9$cm and at time $t=100$s, for the aligned (left) and for the misaligned flow (right).}
	\label{fig:dens_comp}
	\end{figure*}

The flow moves upwards without deviating from the initial direction and as a uniform cylinder. Since its thermal pressure is much greater than that of the surrounding corona, the flow expands at the tip and pushes apart the magnetic field where it propagates. However, since the magnetic field is tightly anchored at the footpoint of the flux tube, its expansion rapidly stops (with Alfv\'enic time scales) and the tension constrains back the flow. As a result, a thin shell of overdense plasma forms all around the head of the flow, as shown in Fig.~\ref{fig:comp_vert} and Fig.~\ref{fig:dens_comp}, which is progressively smoothed out. Overall, the flow keeps its uniformity and cylindrical symmetry while it propagates upwards.

\subsection{The misaligned flow}
\label{subsec:shockdyn}

\begin{figure*}[!th]
	\centering
	\resizebox{\textwidth}{!}
	{
	\includegraphics[width=7.5cm,angle=-90]{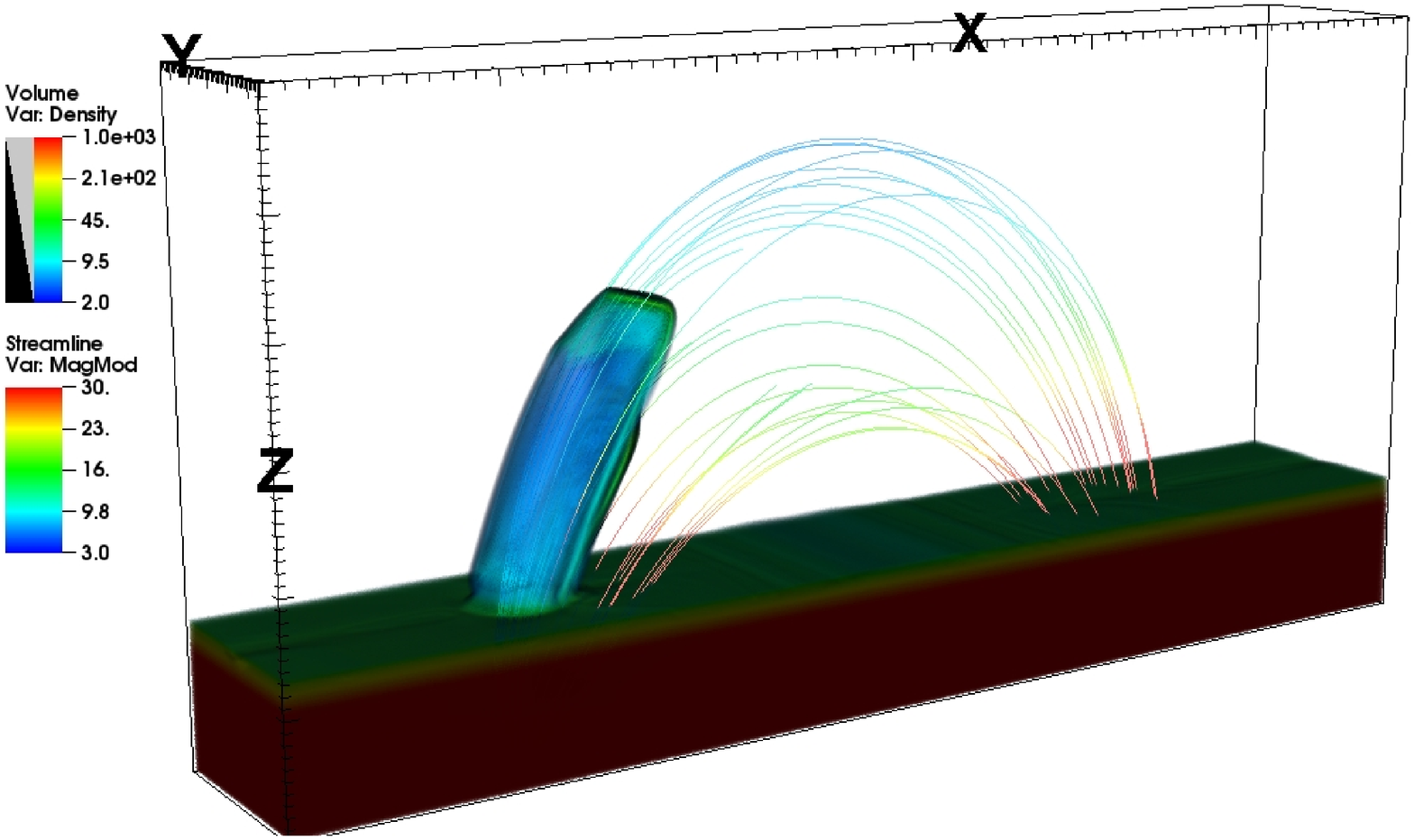}
	\includegraphics[width=7.5cm,angle=-90]{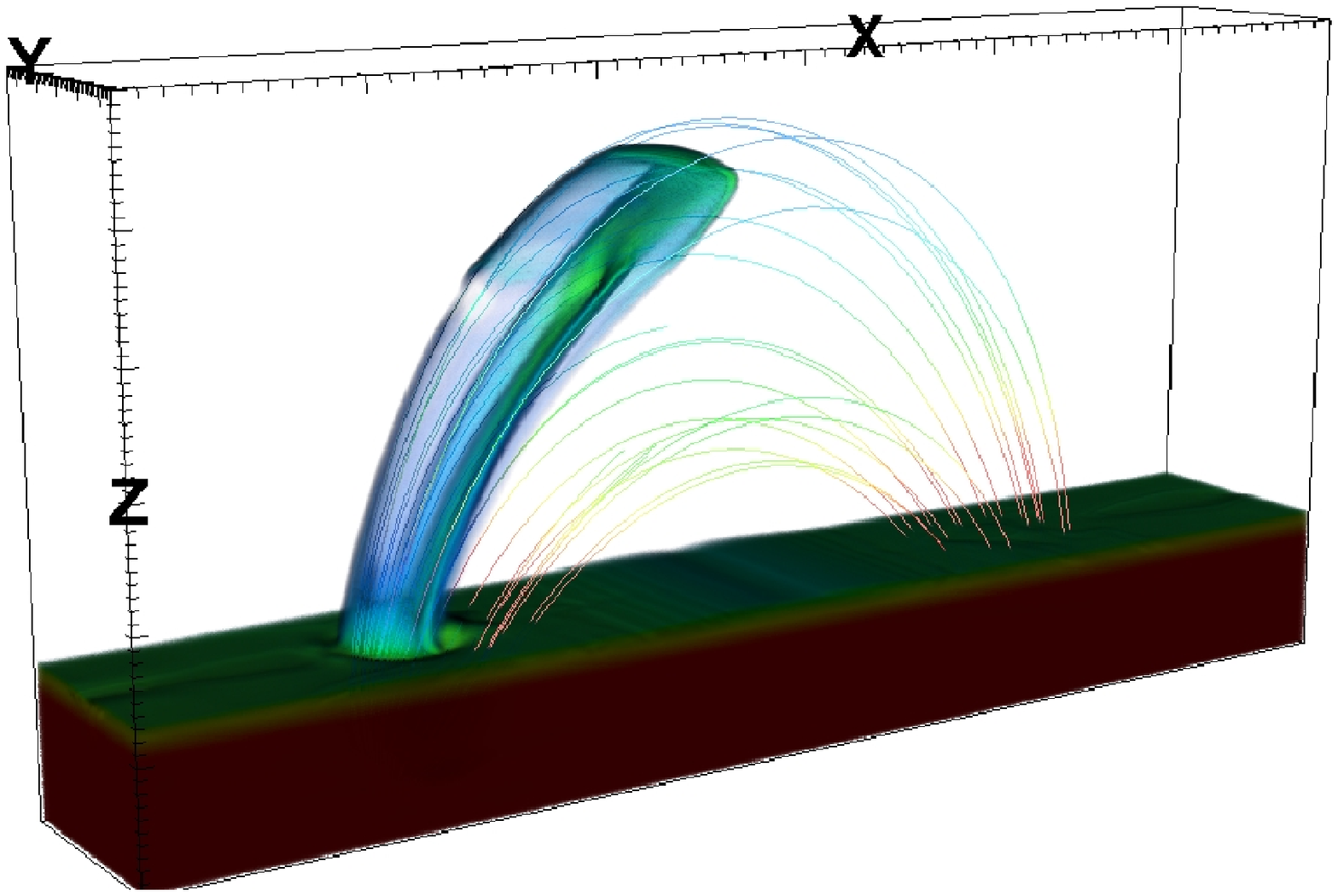}
	}
	\resizebox{\textwidth}{!}
	{
	\includegraphics[width=7.5cm ,angle=-90]{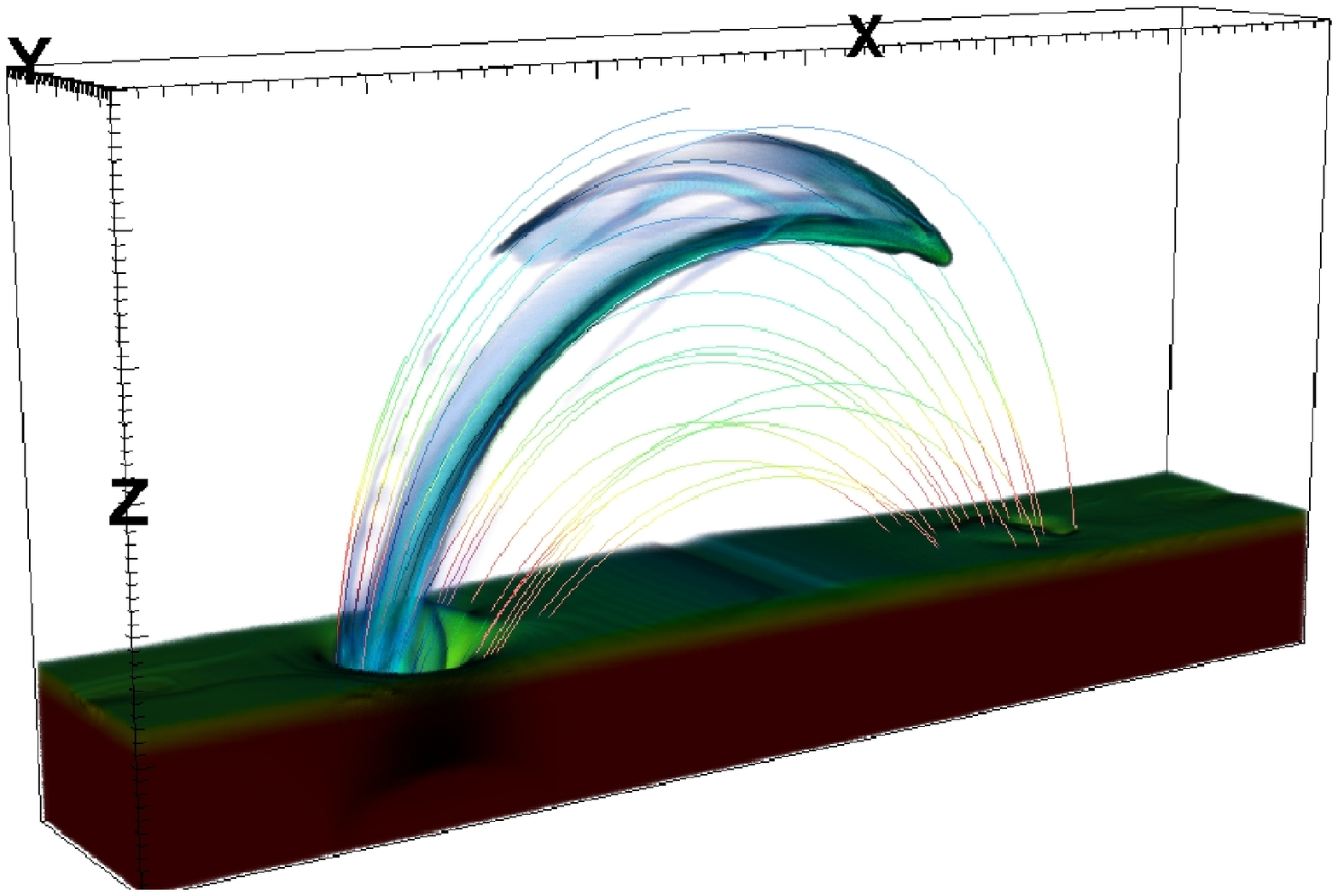} 
	\includegraphics[width=7.5cm ,angle=-90]{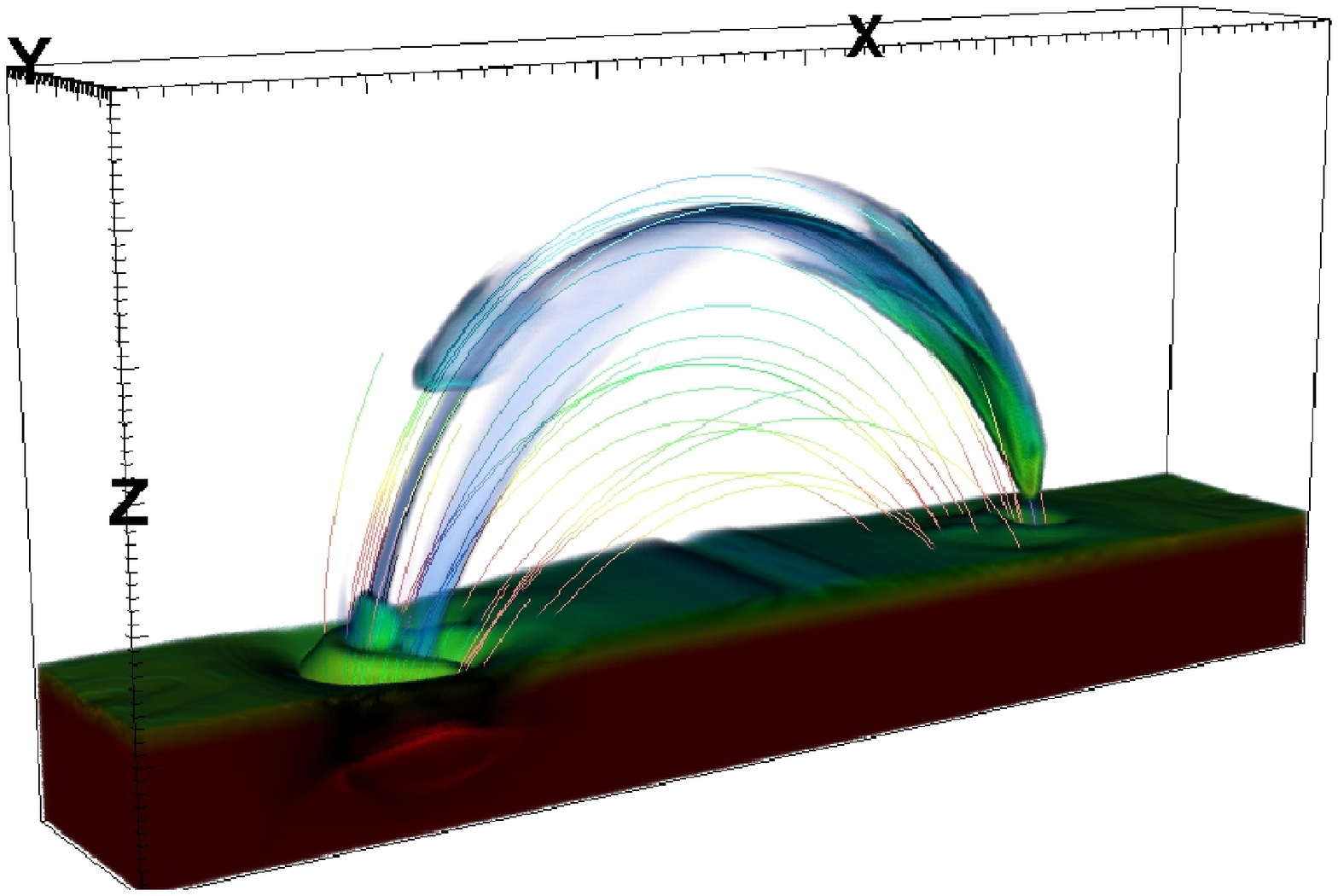} 
	}	
	\caption{Snapshots from our simulation of a misaligned flow: Rendering of the density ($10^{9}$cm$^{-3}$, logarithmic scale) at times t=80, 160, 320, 480s in which a bundle of magnetic field lines \new{(Gauss)} is shown \new{(see the associated movie)}.}
	\label{fig:evol3D}
	\end{figure*}

Figure~\ref{fig:evol3D} shows some snapshots of the evolution of the misaligned flow (see also the associated movie). Although overall the flow still moves along the magnetic flux tube, the detailed evolution is very different from the one of the strictly aligned flow. Again a shock (not visible) is generated and moves ahead of the flow along the tube and the injection is quenched from below. \new{Both the compression ratio of density (~1.7) and the temperature increase (~10\%) are smaller than in the aligned case. This is mainly due to the misalignment, as only the component of the velocity along the field lines contributes. The shock propagates in the hot ambient corona and is potentially visible in the EUV band. Although this is the case closer to the observation described in Section~\ref{sec:obs}, we are actually unable to detect the shock in the data, probably because it is not dense enough to stand out in the thick bright solar limb.}

Since the magnetic flux tube is closed on the chromosphere, the shock hits the other footpoint, at t=$260$~s (when the flow has just crossed the apex of the magnetic channel). Its average speed is therefore $v_{sh}\sim220$km/s.

The bulk plasma motion visible in Figure~\ref{fig:evol3D} and the associated movie has a small component perpendicular to the magnetic field, already since the beginning. Although the plasma is anyway forced to move along the field lines, its misaligned momentum distorts the field above the chromosphere. The flux tube still expands, but the expansion is no longer symmetric.

	\begin{figure*}[!th]
	\centering
	\includegraphics[width=5.5cm, angle=-90]{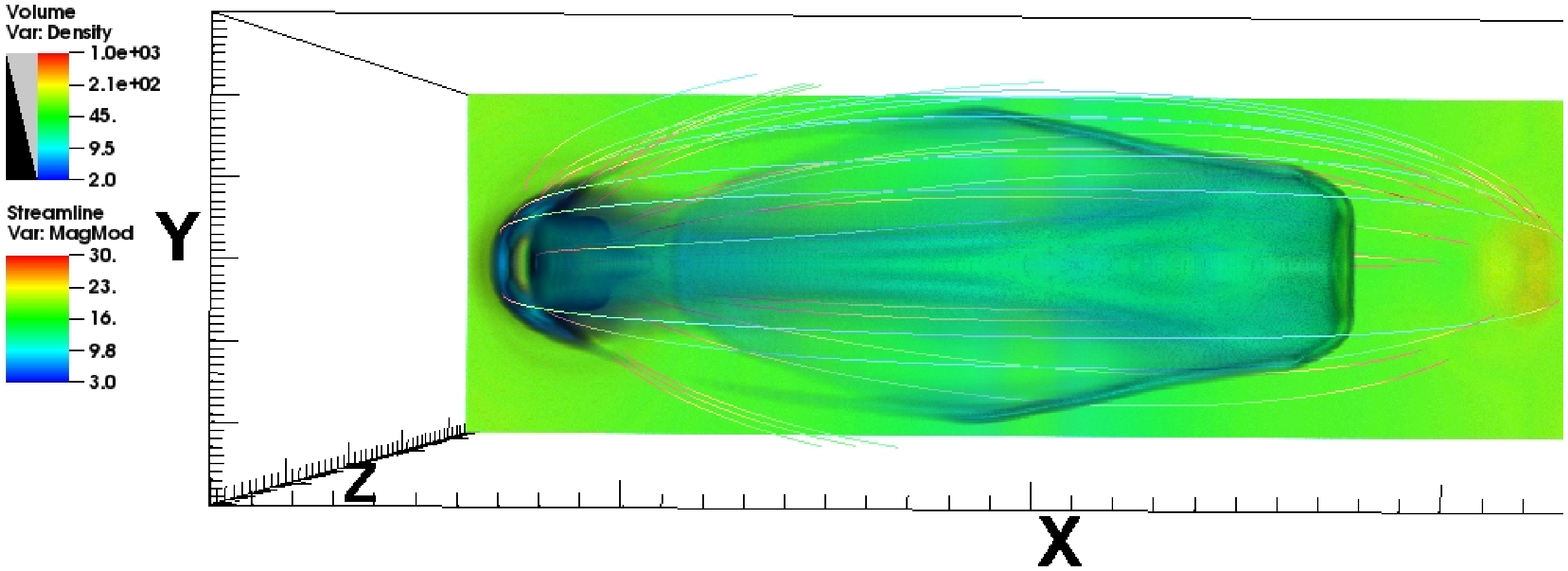}
	\caption{Simulation of the misaligned flow: top view of the density rendering ($10^{9}$cm$^{-3}$, logarithmic scale) at times t=320s in which a bundle of magnetic field lines \new{(Gauss)} is shown.}
	\label{fig:dens_above}
	\end{figure*}

In turn, the deformed field deviates the flow in the direction of the tube curvature, already at $t=20$~s, as we can see in Fig.~\ref{fig:comp_vert}. Since the dense plasma is forced to flow in a direction different from the initial one, its structure becomes asymmetric. It converges and thus is squashed on the side of the curvature, where the magnetic field offers more resistance to deformation. It expands more freely on the other side.

The interplay between the magnetic back-effects (magnetic pressure and tension) and the upward push of the flow is very strong until the flow reaches the apex of the magnetic channel at $t\sim200$~s.
After this time, the magnetic field starts to relax to the initial configuration while the flow falls on the other side of the magnetic channel. The relaxation of the field lines acts as a press that flattens the flux tube and deforms the flow inside it even more. The flow expands laterally at $t\sim320$~s (Fig.~\ref{fig:dens_above}). The relaxation of the field is untidy and frays the flow. Due to this effect, as the flow moves along the tube, it splits becoming highly sub-structured at the end of the evolution at $t\sim480$s (Fig.~\ref{fig:evol3D}).

Overall, we observe an initially cylindrical and uniform flow becoming laminar and filamentary, basically because of the initial misalignment of its direction to the field lines.
 
Fig.~\ref{fig:comp_vert} and Fig.~\ref{fig:dens_comp} compare density maps in different cross-sections at early times and emphasize the difference between the symmetric structure of the aligned flow and the strong bending of the misaligned flow. Fig.~\ref{fig:dens_comp} shows how the thin dense shell all around the aligned flow is squashed all on the side of the curved field in the misaligned flow and eventually determines the laminar structure of the flow.

\section{Discussion and Conclusions}
\label{sec:disc}
In this work, we analyse the dynamics of a continuous flow perfectly or not perfectly aligned to the field lines of a magnetic channel anchored to the solar surface by means of a 3D MHD model analogous to that presented in \cite{Petetal2016}.

Summarizing, we find that the aligned flow travels mostly unperturbed along the magnetic channel, except for some expansion at the tip and for the formation of a dense outer shell. If instead the field lines are strongly curved and the flow is forced to change its direction to follow the magnetic channel, the evolution is significantly different. Because of the stiffness of the field, the flow is squashed to the side of the curved lines and converts into a dense laminar flow, strikingly different from the symmetric dense shell found when the alignment is perfect. As the flow travels through regions of weaker field, it causes some deformation of the field and its untidy back-reaction makes the flow split. The propagation has been monitored until the flow hits the chromosphere at the other side of the closed channel.

We expect that the result can somewhat change for different conditions in the model, but we are unable to quantify without a relevant exploration of the parameter space. For instance, we might expect that a weaker field and/or a more tenuous ambient atmosphere should make the flow expansion easier and perhaps smooth down the dense shell of the aligned flow. A faster outflow in the curved loop should determine a stronger deformation of the field and we might expect an even larger backreaction and flow splitting.

The transformation of the initially cylindrical, symmetric and uniform flow into a strongly laminar and substructured one might explain the puzzling evidence of hedge-like downfalls observed with SDO/AIA (ex. Figure~\ref{fig:obs} and the associated movie). We might be simply in the presence of a relatively long-fed flow that is not launched in the direction of the local magnetic field, probably not particularly strong, and therefore considerably deviated by the field during its journey. 
Our estimate of the flow conditions supports this hypothesis and also gives us an idea of the intensity of the ambient magnetic field, which might be similar to that in our model, e.g. $\sim30$G. This scenario may not be uncommon \citep{Maretal2016}.

The results we found are general and could be applied to many flows moving in a magnetized medium: if the flow motion is not aligned to the magnetic field, the flow tends to be flattened and split by the deviation of the direction of motion. If the alignment is nearly perfect the flow simply travels along the field lines, without significant deformations, as expected for plasma strictly confined in magnetic coronal loops. \new{This effect might determine also the morphology of coronal flows out into the interplanetary space, as those detectable and to be studied by Solar Orbiter instruments.} 

\begin{acknowledgements}
We thank Carolus J. Schrijver for pointing out the SDO observation. AP and FR acknowledge support from the Italian \emph{Ministero dell'Universit\`a e Ricerca}. PLUTO is developed at the Turin Astronomical Observatory in collaboration with the Department of Physics of the Turin University. We acknowledge the CINECA Award HP10B59JKR. The simulations have been partly run on the Pleiades cluster through the computing project SMD-16-7704 from the High End Computing (HEC) division of NASA. P.T. was supported by NASA grants NNX14AI14G, NNX15AF50G and NNX15AF47G. This work has benefited from discussions at the International Space Science Institute (ISSI) meetings on 'New Diagnostics of Particle Acceleration in Solar Coronal Nanoflares from Chromospheric Observations and Modeling' where topics relevant to this work were discussed with other colleagues.
\end{acknowledgements}

\bibliographystyle{aa}
\bibliography{upflows}

\end{document}